# Preparing Undergraduates for Research Careers: Using Astrobites in the Classroom

by **Nathan E. Sanders,**
Harvard University,
**Susanna Kohler,**
University of Colorado Boulder,
**Elisabeth Newton,**
Harvard University,
and the **Astrobites collaboration**

## Abstract

Because undergraduate participation in research is a longstanding and increasingly important aspect of the career path for future scientists, students can benefit from additional resources to introduce them to the culture and process of research.  We suggest the adoption of the web resource *Astrobites* as a classroom tool to increase the preparation of undergraduate physics and astronomy students for careers in research.  We describe the content and development of the website, discuss previous university courses that have made use of *Astrobites*, and suggest additional strategies for using *Astrobites* in the classroom.

## 1. Introduction

Research experience is a vital aspect of undergraduate education in the physical sciences, as well as an increasingly-critical qualification for graduate admissions (Crowe and Brakke 2008, Russell et al. 2007, DeHaan 2005, Wilkerson 2007).  However, many developmental barriers stand in the way of undergraduate students' successful participation in research, even for those with strong conceptual and technical backgrounds from coursework. Comprehension of the specialized jargon and idioms of the field, understanding of research practices and technical methods, and familiarity with the technical literature are useful knowledge that students new to research lack. The challenge of obtaining this background can be particularly intimidating to undergraduates working with more senior researchers who acquired a more complete command of these faculties through a decade or more of undergraduate, graduate, and post-doctoral study.

Immersion in scientific discourse through activities such as seminar attendance and interaction with senior researchers has been a leading pathway for undergraduates to attain literacy in research, similar to the widely-applied philosophy for education in foreign language (Johnson and Swain 1997).  However, the time available for such direct interaction is necessarily limited, and can vary substantially depending on the size and structure of the host department at the student's institution.  To supplement these important real-world interactions, virtual tools can be used to increase student immersion (Price 2007, Chang et al. 2012).

In this commentary, we suggest the use of the web resource *Astrobites* as a learning tool for undergraduate astronomy courses.  We describe the content and goals of the *Astrobites* website and its authorship, discuss previous uses of *Astrobites* in formal academic settings, and suggest novel applications of its content to prepare undergraduates for careers in research.

## 2. Describing Astrobites

**Figure 1.** Screenshot of the Astrobites website, featuring the summary of Schneider et al. (2012). Accessed 2012 April 30.

The *Astrobites* website (http://astrobites.com), founded in December 2010, is an online "reader's digest" for the astrophysics preprint server *astro-ph*.[1] As illustrated in Figure 1, the website publishes succinct (~4–7 paragraph) summaries of the methods, results, conclusions, and context of new research papers in astrophysics posted to astro-ph. The summaries are designed for undergraduates who have taken coursework in the physical sciences. Knowledge of basic physical concepts (e.g. energy conservation or blackbody radiation) is assumed, but relevant terminology and research methods are defined and links are provided to detailed references. Astrobites articles incorporate figures from the original sources, guiding readers through the visual presentation of key results from the paper. About 400 of these summaries have been posted to date, with a current publication rate of one per day. *Astrobites* also publishes pieces with career advice for young researchers, notices of professional opportunities available to undergraduate students (such as Research Experiences for Undergraduates, REU, programs), and other supplementary content at a rate of twice per week.

*Astrobites* was created and is developed and maintained by graduate students in astrophysical sciences from the United States and Europe (including the authors of this commentary). A regular "rotation" of about 30 authors each write 1 *Astrobites* post per month and an additional set of about 10 graduate students provide supplemental contributions.[2] The motivation for drawing on graduate students to prepare content for *Astrobites* is two-fold: 1) researchers at this stage of their career are sufficiently expert in the content of the research papers to prepare summaries, but are also aware of the level of familiarity to expect from the undergraduate target audience, and 2) the graduate students benefit from the pedagogical exercise of summarizing research papers into an accessible format.

---

[1]        astro-ph is the Astrophysics section of the arXiv, http://arxiv.org/archive/astro-ph

[2]        The full *Astrobites* authorship is listed at http://astrobites.com/meet-the-authors/   All *Astrobites* authors contribute on an unpaid basis and the website is presented without advertising and entirely as a not-for-profit enterprise.

*Astrobites* currently receives about 300 regular readers per day, as measured by returning visitors tracked by Google Analytics.  In addition, an informal September 2011 reader survey (for which there were 125 respondents) indicated that 2/3 of our readers do not visit our website, but instead subscribe to our RSS feed.  The survey also showed that 20% of the respondents were undergraduate students.  The remainder of the readership is approximately evenly divided between graduate students in astronomy, researchers who have already earned the PhD, and non-professional astronomy "enthusiasts."  We therefore estimate that *Astrobites*' daily readership already includes >150 students in our target demographic of undergraduates in a physical science field.  In the same reader survey, >80% of undergraduate respondents listed their career goal as "researcher" in astrophysics or a related field.

# 3. Astrobites in the Classroom

## 3.1 Past uses of Astrobites

To assess the usage of *Astrobites* in undergraduate classrooms to date, we performed a survey targeted at both students and educators in our readership that have used *Astrobites* in the classroom.[3]  We received responses from 4 students and 6 educators indicating that they had used *Astrobites* in a class.  From among these responses, we present two anecdotes regarding the use of Astrobites in undergraduate classrooms:

1. **Astronomy 20, " Basic Astronomy & the Galaxy"**
   California Institute of Technology, Fall 2012, Prof. John Johnson

   In this inquiry-based introductory astronomy course, students were instructed to maintain their own blogs.[4]  Students used the blogs to record problem-solving techniques learned in the class, as well as to produce original content including interviews with astronomers and summaries of modern research topics.  *Astrobites* was presented to the students as a model for a blog of this type.

2. **Astronomy 61, "Current Problems in Astronomy and Astrophysics"**
   Swarthmore College, Spring 2012, Prof. Eric Jensen

   In this journal-club-style discussion course for advanced undergraduates, students were responsible for selecting and summarizing research papers for the week's seminar on a rotating basis.  *Astrobites* was offered to students as a resource for selecting interesting papers and as a source for background information and context.

These anecdotes illustrate a subset of the possibilities for incorporating Astrobites into formal educational settings to increase student engagement outside the classroom.

## 3.2 Suggested uses of Astrobites

---

[3]     The *Astrobites* educator survey questionnaire is available at http://astrobites.com/2012/03/05/aersurvey/

[4]     The student blogs from the Astrobomy 20 course can be accessed at http://www.astro.caltech.edu/~jrv/Ay20/Ay20_Class_Website/Blogs.html

To supplement the existing examples of *Astrobites* usage in the classroom described in the previous section, here we present suggestions for novel applications of *Astrobites* in undergraduate curricula.

1.  **Reading assignments**: Students are asked to read *Astrobites* throughout the semester, to increase immersion in astronomy outside the classroom and to reinforce concepts learned in lectures. Astrobites would therefore be used in a manner similar to web-based multimedia learning modules as pre-lecture assignments to supplement textbook materials (Stelzer et al. 2009, Sadaghiani 2012). Student participation could be evaluated during in-class discussions of current research, or by asking students to formulate follow-up questions based on the reading. These questions could be submitted for class discussion, or could be posted directly on the *Astrobites* website (see 4 below).

2.  **Written summaries:** Students are asked to write their own summaries of research papers in astronomy, modeled on the format of *Astrobites*. These research summaries could be assigned periodically throughout the semester, or as a single term project where the students are asked to review several topical papers. The Astrobites website provides a convenient archive of brief, informative, and accessible summaries of recent research in astrophysics and can therefore serve as a resource to help students identify specific studies or topics that have methodology and results that interest them. This obviates the need for students to become sufficiently expert in the full range of astrophysics research to select a recent paper from a general bibliography of publications. If student writing is posted online, in a blog format, it may facilitate peer discussion and encourage students to make high-order connections between scientific topics discussed in the classroom and research methods (Daniels 2010).

3.  **In-class presentations:** Students are asked to describe a research paper in astrophysics to their peers via an in-class presentation. As in #2 above, the Astrobites website provides an archive of summaries which students can use to select interesting papers or topics. Moreover, if students plan to present on a paper that was already discussed on Astrobites, their classmates can read the Astrobites summary to rapidly familiarize themselves with the subject matter so they may come prepared to start a discussion. In this way, students can engage in a collaborative argumentation learning environment where they can debate the merits of the scientific ideas presented in the papers, as well as the assumptions that the ideas rely on (Bell and Linn 2000).

4.  **Discussion forum:** Students are asked to use the commenting feature of the *Astrobites* website to hold a discussion forum on a given research topic. By holding this discussion on the website rather than in the classroom, students have the opportunity to formulate their responses after doing background research into the topics as necessary and to cite relevant materials. Furthermore, because Astrobites authors and researchers from throughout the world (including, often, the authors of the paper in discussion) read Astrobites, the discussion may broaden to include a wide range of perspectives and expertise.[5] Comments from readers at all levels of instruction would be welcome on the Astrobites website.

# 4. Conclusions

---

[5] Examples of Astrobites articles with reader discussions can be found at the following links:
http://astrobites.com/2011/07/13/cold-flows-and-the-first-quasars/
http://astrobites.com/2011/04/24/inelastic-dark-matter-ruled-out/

In this commentary, we have presented *Astrobites* as a web resource that can be integrated into formal classroom education to improve undergraduate preparation for careers in research. We have described previous applications of *Astrobites* to the undergraduate curriculum and suggested novel uses for the website in future curricula.

In the sense that *Astrobites* is written by and for students in astronomy, and because many of the strategies we have suggested involve classroom discussions and peer-teaching, the usage of *Astrobites* in the classroom is a form of peer-assisted learning (see e.g. Parkinson 2009). If the adoption of *Astrobites* as a mechanism for this type of instruction becomes more widespread, it could facilitate the use of statistically-validated studies on the educational impact of *Astrobites* or other web-based pedagogy tools in the classroom.

## Acknowledgements

We thank the anonymous referee for helpful comments. We are grateful to John Johnson, Eric Jensen, and all the other respondents to our survey for sharing their teaching experiences and to Douglas Duncan, Thomas Hockey, and Phil Sadler for their helpful discussions and support. Web hosting for *Astrobites* is provided by James Guillochon at the University of California at Santa Cruz. N.E.S. and E.N. are supported by the National Science Foundation through Graduate Research Fellowships, and S.K. is supported by NSF grant AST-0907872.